\documentclass{ws-procs9x6}

\begin{document}

\title{Background studies \\ 
for a ton-scale argon dark matter detector (ArDM)\footnote{Invited talk at the
6th International Workshop on the Identification of Dark Matter, September 2006,
Island of Rhodes, Greece.}}

\author{L. Kaufmann and A. Rubbia}

\address{Institute for Particle Physics, ETH Zurich, Switzerland \\
E-mail: lilian.kaufmann@cern.ch, andre.rubbia@cern.ch}

\begin{abstract}
The ArDM project aims at operating a large noble liquid detector to 
search for direct evidence of Weakly Interacting Massive Particles 
(WIMP) as Dark Matter in the universe.
Background sources relevant to ton-scale liquid and gaseous argon 
detectors, such as neutrons from detector components, muon-induced 
neutrons and neutrons caused by radioactivity of rock, as well as 
the internal $^{39}Ar$ background, are studied with simulations.
These background radiations are addressed with the design of an 
appropriate shielding as well as with different background 
rejection potentialities. Among them the project relies on event 
topology recognition, event localization, density ionization 
discrimination and pulse shape discrimination.
Background rates, energy spectra, characteristics of the background-induced 
nuclear recoils in liquid argon, as well as the shielding 
performance and rejection performance of the detector are described.
\end{abstract}

\bodymatter

\section{Introduction}

Astronomical observations suggest that over 80\% of the matter contained 
in the universe is Dark Matter, which presumably consists of unknown and 
invisible WIMPs (Weakly Interacting Massive 
Particles). The most common candidate is the lightest supersymmetric 
particle. Direct detection experiments aim at 
measuring elastic scattering of WIMPs with nuclei.
The energy transferred to the nuclei is typically below 100 keV. Due 
to this low energy range and the small cross section, such a signal is 
difficult to measure and signal sensitivity is limited by ordinary backgrounds. 
Background radiation arises from several sources and is of different 
significance. One dominant background 
source is the contamination
with radioactive elements, namely uranium and thorium chains, and potassium-40
resulting in $\alpha$, $\beta$ and $\gamma$ emissions as well as 
neutron production processes, of the materials used in and around the detector. 
Another source of neutron background in
underground environments comes
from interactions of cosmic muons with matter, resulting in neutron 
production. A third background source can arise from internal contamination 
of the target material, e.\,g.~with radioactive isotopes. \\
The experiment for which the present study has been carried 
out is ArDM~\cite{IC1,Kaufmann:2006hp,Rubbia:2005ge,Talk}, a ton-scale liquid argon detector. The technical concept of the experiment 
relies on the independent readout of ionization charge by LEM and 
of scintillation light by photo-detectors~\cite{Rubbia:2005ge}. This paper summarizes 
the results of Monte Carlo studies carried out to assess the impact of different 
background radiation in this project.

\section{Neutron background} 

An important background is neutron-induced
nuclear recoils which are 
hardly distinguable from WIMP events. 
The neutrons from radioactivity arise mainly from contaminations of materials or rock 
by the two radioactive elements uranium and thorium, and their decay-chain daughters.  
The decay chains of $^{238}$U 
and $^{232}$Th contain $\alpha$ decays with $\alpha$ energies of 3.5 to 11 MeV. 
These $\alpha$ undergo ($\alpha$,n) reactions, thereby producing neutrons 
with energies in the MeV range. The cross section depends on the material 
and on the $\alpha$-energy \cite{Heaton}. Typical yields are 10$^{-8}$ to 
10$^{-5}$ neutrons per $\alpha$. 
Besides ($\alpha$,n) reactions, neutrons can arise from spontaneous fission. The number 
of neutrons arising from these two processes is typically of the same order of magnitude, 
but depends on the material. 
Muon-induced neutrons arise from cosmic muon interactions with surrounding
materials. Highly energetic muons 
are able to penetrate deep underground. 
Neutrons are produced by spallation or photonuclear
processes, or by secondary interactions
of muon-induced hadronic showers. 
The shielding of the detector optimized for natural radioactivity
is less efficient in this case and faster neutrons can penetrate
inside the detector fiducial region.
Shielding and detector components can also act 
as a target for muons, however the expected production rates are low. 

\subsection{Neutrons from radioactivity in detector components}

The ArDM detector parts consist mainly of the following materials: stainless steel, vetronite, 
polyethylene, borosilicate glass, ceramics and copper (see Ref.~\cite{Rubbia:2005ge} for details). 
In general, metallic materials contain much less U/Th contamination, i.e. typically a few ppb. 
Minerals and glasses generally have higher contaminations of the order of a few hundred ppb. 
We have estimated the number of emitted neutrons using the data from Ref.~\cite{Heaton}. Furthermore, 
the emission numbers have been simulated with SOURCES4a for cross-check. The numbers obtained 
with the two different approaches agree well.
The biggest contribution for standard materials comes from photomultiplier tubes located on 
the bottom of the detector and large electron multiplier (LEM) plates located on the top of the detector, 
since these two parts contain glass. 
Low background versions of these two components should
reduce the rate by about two orders of magnitude. 
The resulting neutron production rates and the precise assumptions for component masses and contaminations 
are summarized in Table~\ref{tab:mass}.

\begin{table}
\begin{center}
\tbl{Assumptions for detector components and estimated neutron production based on \cite{Heaton} and SOURCES simulation. A description of the detector geometry can be found in \cite{Kaufmann:2006hp,Talk, Rubbia:2005ge}.}
{\begin{tabular}{|l|l|l|l|l|}
\hline
Component & Mass (kg) & ppb U & ppb Th & n/year \\ 
\hline
\hline
Dewar (steel)	 & 1000 & 0.6 & 0.7 & 380 \\
\hline
LEM Vetronite & 2  & 1000 & 1000  & 10000 \\
\hline
LEM	low bg. & 4  & $<$2 & $<$2 & $<$40\\
\hline
PMT std. version & 2.4 (80 tubes) & 400 & 400 & 12000  \\
\hline
PMT low bg. & 9.8 (14 tubes) & 30 & 30  & 1400 \\
\hline
Pillars	& 13 (8 Pillars) & 20 & 20  & 310 \\
\hline
\end{tabular}}
\label{tab:mass}
\end{center}
\end{table}

\subsubsection{Neutron spectra}

\begin{figure}
\begin{center}
\begin{tabular}{l l}
\hspace{-0.6 cm}
\includegraphics[width=0.53\textwidth]{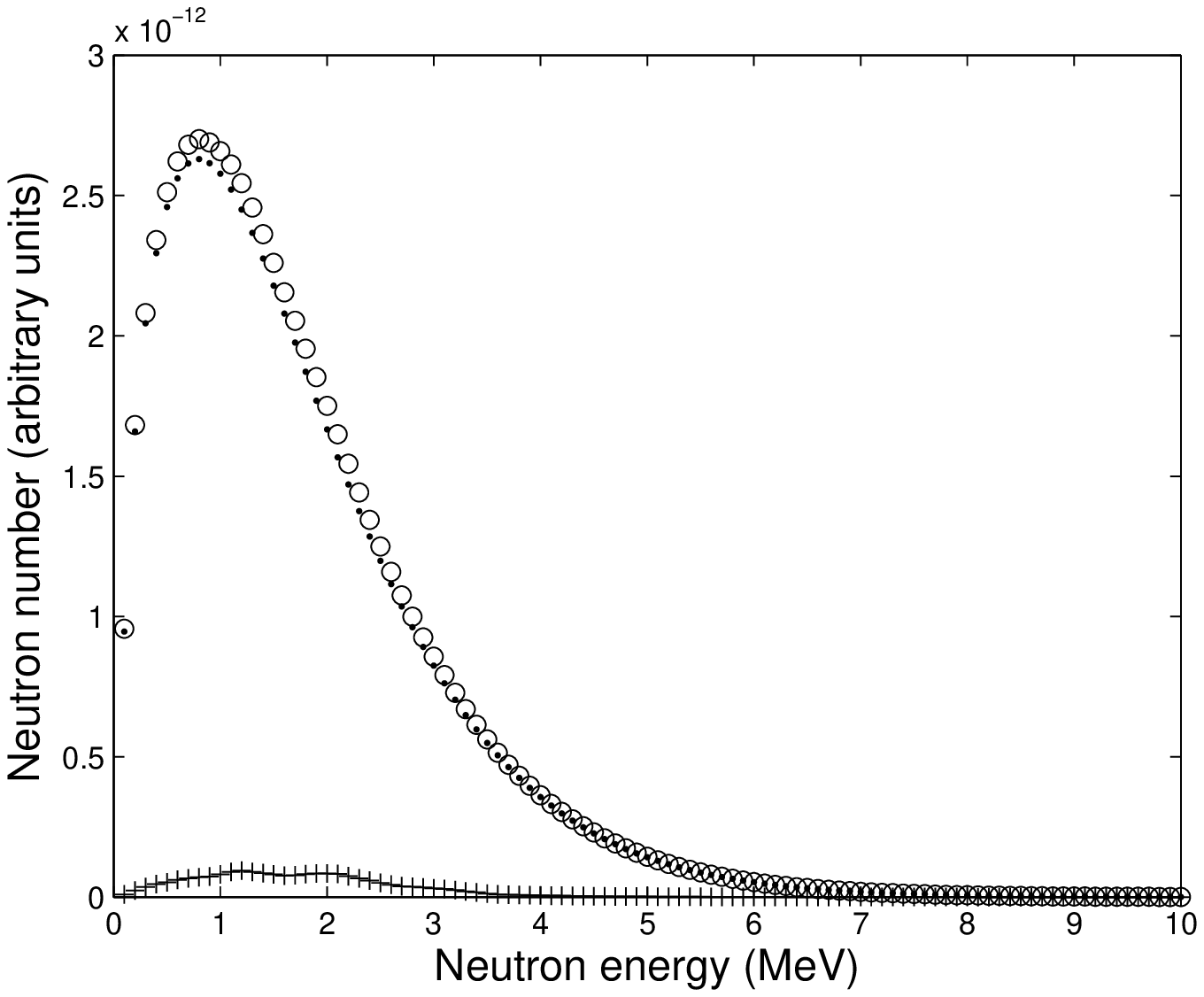} & \hspace{-0.8cm} 
\includegraphics[width=0.53\textwidth]{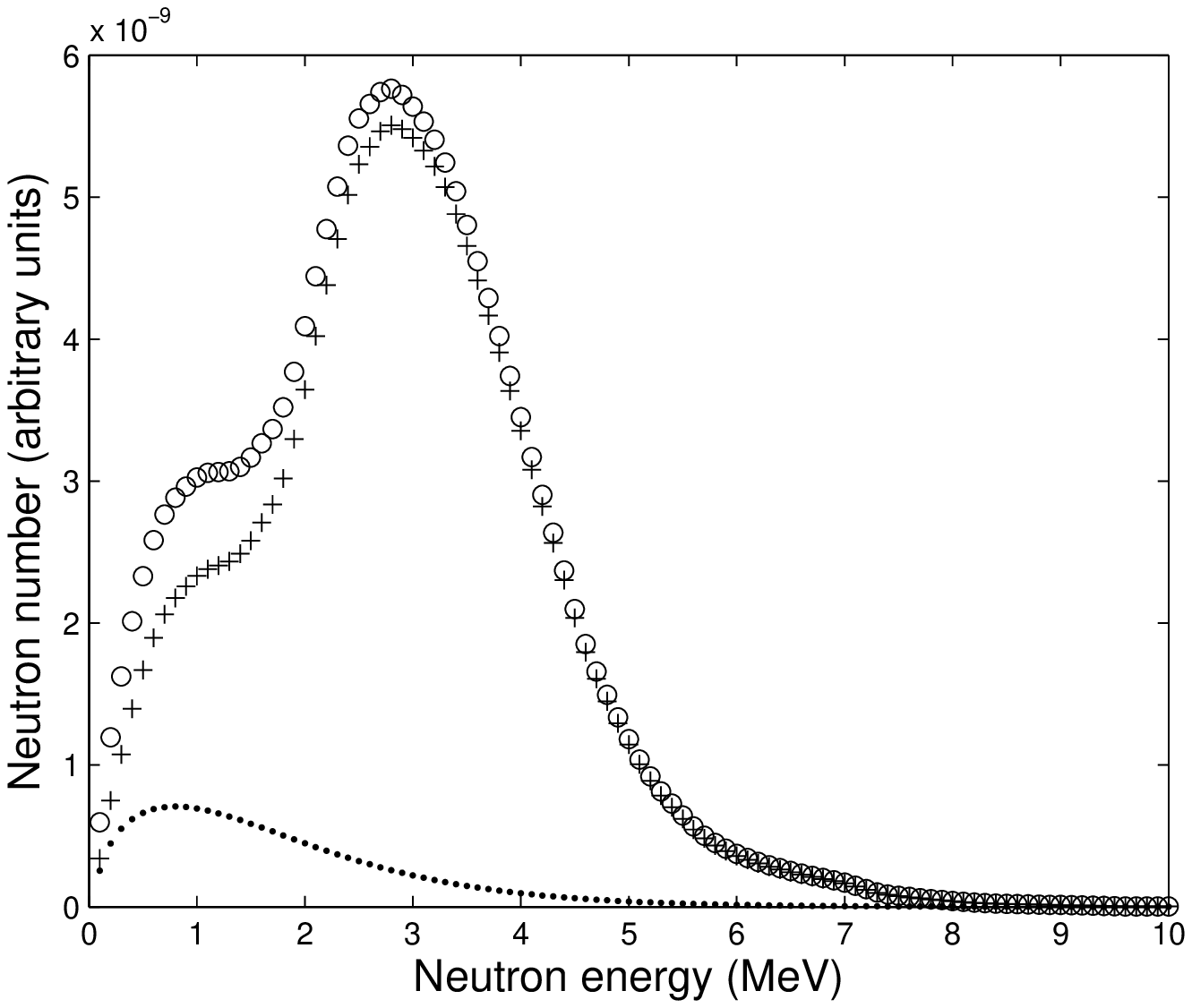} \\
\end{tabular}
\caption{Simulated neutron energy spectra for the dewar (left) and LEM (right). The dotted line is the contribution from spontaneous fission, the crossed line the contribution from ($\alpha$,n.) reactions and the circled line is the sum of the two.}
\label{sources}
\end{center}
\end{figure}

The energy spectra of neutrons coming from U and Th decay chains have two contributions, 
namely the one from spontaneous fission and the other from ($\alpha$,n) reactions. 
The spontaneous fission spectrum is described by $dN/dE\propto\sqrt{E}\mathrm{exp}(-E/1.29) $.
About 2 neutrons are emitted per spontaneous fission. 
The spectrum of neutrons coming from ($\alpha$,n) reactions is more 
involved, since the ($\alpha$,n) cross section is 
material-dependent.   
For the computation of the spectra in detector component materials, the simulation program SOURCES4a 
was used. The code was extended to $\alpha$ energies above 6 MeV by \cite{Carson}. 
The resulting spectra for stainless steel (dewar) and Vetronite (LEM) are shown in Fig.~\ref{sources}. 

\subsection{Neutrons from radioactivity in rock and concrete}

The minerals constituting the rock overburden in an underground laboratory also contain small amounts 
of U and Th, causing neutron radiation in the same way as in the detector components. 
The level of the contamination strongly depends on the location and the elemental composition of the rock. 
Usually the U/Th contamination is on the order of a few hundred ppb. 
The studies described here have been carried out for the Canfranc Underground Laboratory. It is located in the Pyrenees at a depth of 2450~m.w.e. 
A concrete layer covers the walls of the laboratory, which also affects the 
neutron radiation flux. \\
CH$_2$ is the typical material used as shield against these ambient neutrons.  
The number of neutrons coming from rock and concrete and reaching different depths of the shielding was simulated with GEANT-4~\cite{Agostinelli:2002hh}. The neutron flux decreases by approximately one order of magnitude after every 10 cm of CH$_2$ at energies 
up to 6 MeV, for higher energies the suppression is slightly less.  The suppression for concrete neutrons 
is slightly lower, expecially for higher energies. On the average, 60 cm of CH$_2$ reduce the flux of neutrons by about six orders of magnitude.

\subsection{Detector response to neutrons} 

\begin{figure}
\begin{center}
\includegraphics[width=0.5\textwidth]{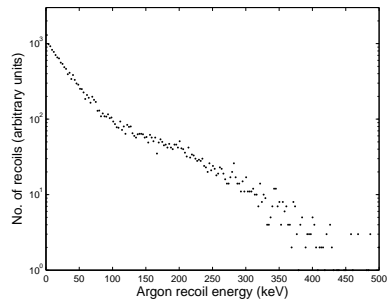}
\caption{Spectrum of argon recoils caused by neutrons from detector components.}
\label{recoilspectrum}
\end{center}
\end{figure}

Background neutrons interact with the detector via elastic scattering with argon nuclei. The imparted recoil 
energy on the nucleus caused by a neutron with energy $E_n$ and a scattering angle $\theta$ is $E_R\simeq 2E_n\frac{M_n\,M_{Ar}}{(M_n+M_{Ar})^2}(1-\mathrm{cos}\theta) $.  
Neutrons from the various sources mentioned above have been simulated with the
proper energy spectrum and fully propagated through a detailed geometry of
the detector with GEANT-4. As an example,
the resulting argon recoil spectrum for neutrons from detector components is shown 
in Fig.~\ref{recoilspectrum}. 

\noindent
Of the neutrons entering the fiducial volume, more than 50\% can be rejected because of multiple scattering, since a WIMP would never scatter more than once within the detector active volume. 
The number of neutron scatters depends on the size of the fiducial volume, the distance of the neutron emitting components to the fiducial volume and on the energy threshold. Assuming that 
a WIMP-like event is a single recoil with an
energy between 30 keV and 100 keV, approximately 5\% of the neutrons from detector components produce WIMP-like events. The number of remaining events per year is 12 for the neutrons from the dewar, 800 for the standard material LEM, below 2 for the low background LEM, 480 for the standard PMTs, 56 for the low background PMTs and 19 for the pillars; to be compared with 3500 WIMP events at $\sigma_{\mathrm{WIMP-nucleon}}=10^{-7}$ pb. 

\section{$^{39}$Ar electron background}
\label{Ar39}

Commercially available Argon is procured by liquefaction of air and
contains radioactive isotopes. $^{39}$Ar decays via $\beta$-disintegration into $^{39}$K with a 
half-life of 269 years and Q=565 keV. 
The concentration of $^{39}$Ar in atmospheric argon is (7.9$\pm0.3)\cdot$ 
10$^{-16}$g/g \cite{Loosli}, causing a decay rate of 1 kHz in one ton of argon.
$\gamma$s from U/Th of the detector 
components produce an interaction rate which is about three orders of magnitude smaller. \\
The rejection of electron and $\gamma$ events is facilitated by two means: charge/light ratio discrimination~\cite{IC1} and 
pulse shape discrimination~\cite{Boulay:2006mb}. 
The first uses the fact that the ionization yield of nuclear recoils is highly
quenched compared to that of electron/$\gamma$, while the scintillation yield is similar. 
The pulse shape analysis relies on different populations of the fast and slow
components of scintillation.  
In order to overcome the internal $^{39}$Ar background, a combined rejection power 
of about 10$^8$---depending on the WIMP-nucleon cross section---is required.  

\section{Conclusion}

Three main background sources for the ArDM experiment---a liquid argon ton-scale detector---have been studied. 
Neutron radiation is the most important background, since neutron events have the same signature inside the detector as WIMP events. The main neutron background comes from contaminations of detector components with radioactive 
elements. 
Neutrons from rock and concrete of the underground laboratory can be shielded by a thick enough CH$_2$ shield. 
The internal $^{39}$Ar background is strongly suppressed if the light/charge ratio and the scintillation light time distribution are measured precisely enough.

\end{document}